\RequirePackage{fix-cm} 
\documentclass[a4paper, twoside, reqno, dvips, 12pt]{amsart}
\usepackage{fixltx2e}   

\usepackage[latin1]{inputenc}
\usepackage[T1]{fontenc}

\usepackage{eucal}
\usepackage{esint}
\usepackage{xspace}
\usepackage{amsgen}
\usepackage{amssymb}
\usepackage{amsmath}
\usepackage{amsfonts}
\usepackage{MnSymbol}
\usepackage[nice]{nicefrac}
\usepackage{mathrsfs, dsfont}

\usepackage{a4wide}

\usepackage[table]{xcolor}

\definecolor{oneblue}{rgb}{0.0, 0.0, 0.85}
\definecolor{darkgrey}{rgb}{0.273, 0.281, 0.30}
\definecolor{Lightgray}{rgb}{0.89, 0.89, 0.89}
\definecolor{Lightblue}{RGB}{214, 214, 214}
\definecolor{bckg}{RGB}{20.8, 20.8, 20.8} 

\usepackage{psfrag}
\usepackage{graphicx}
\usepackage{subfigure}
\usepackage{morefloats}
\usepackage{indentfirst}

\usepackage{acronym}
\usepackage{microtype}
\usepackage[labelfont={bf,sf},%
            labelsep=period,%
            justification=raggedright]{caption}

\usepackage[usenames, dvipsnames, pdf]{pstricks}
\usepackage{epsfig}
\usepackage{pst-grad} 
\usepackage{pst-plot} 

\usepackage[colorlinks,
            urlcolor=oneblue,
            linkcolor=oneblue,
            citecolor=oneblue,
            bookmarksopen=false,
            pagebackref]{hyperref}

\usepackage[explicit]{titlesec}

\titleformat{\section}
  {\color{darkgrey}\Large\sffamily\bfseries}
  {}
  {0em}
  {\colorbox{bckg!5}{\parbox{\dimexpr\linewidth-2\fboxsep\relax}{\centering\thesection. #1}}}
  []

\titleformat{name=\section,numberless}
  {\color{darkgrey}\Large\sffamily\bfseries}
  {}
  {0em}
  {\colorbox{bckg!10}{\parbox{\dimexpr\linewidth-2\fboxsep\relax}{\centering#1}}}
  []

\titleformat{\subsection}
  {\color{darkgrey}\large\sffamily\bfseries}
  {}
  {0em}
  {\colorbox{bckg!5}{\parbox{\dimexpr\linewidth-2\fboxsep\relax}{\centering\thesubsection. #1}}}
  []

\titleformat{name=\subsection,numberless}
  {\color{darkgrey}\Large\sffamily\bfseries}
  {}
  {0em}
  {\colorbox{bckg!10}{\parbox{\dimexpr\linewidth-2\fboxsep\relax}{\centering#1}}}
  []

\titleformat{\subsubsection}
  {\sffamily\normalsize\bfseries}
  {\thesubsubsection}
  {0em}
  {#1}
  []

\titleformat{\paragraph}[runin]
  {\sffamily\small\bfseries}
  {}
  {0em}
  {#1}

\titlespacing*{\section}{1.0em}{1.0em}{0.8em}[0em]
\titlespacing*{\subsection}{1.0em}{1.0em}{0.8em}[0em]
\titlespacing*{\subsubsection}{1.0em}{0.7em}{0.6em}[0em]

\usepackage{titletoc}

\contentsmargin{0.55em}

\newlength{\tocsep}
\titlecontents{section}[\tocsep]
  {\addvspace{10pt}\bfseries\sffamily}
  {\contentslabel[\thecontentslabel]{\tocsep}}
  {}
  {\ \titlerule*[0.75pc]{.}\ \thecontentspage}
  []
\titlecontents{subsection}[\tocsep]
  {\addvspace{8pt}\sffamily}
  {\contentslabel[\thecontentslabel]{\tocsep}}
  {}
  {\ \titlerule*[0.5pc]{.}\ \thecontentspage}
  []
\titlecontents{subsubsection}[\tocsep]
  {\addvspace{7pt}\footnotesize\sffamily}
  {\contentslabel[\thecontentslabel]{\tocsep}}
  {}
  {\ \titlerule*[0.5pc]{.}\ \thecontentspage}
  []

\usepackage{fancyhdr}
\usepackage{lastpage}

\newcommand*\Title{Resonance enhancement by detuning}
\newcommand*\Authors{D.~Dutykh \& E.~Tobisch}
\newcommand*{\plogo}{{\texttt{arXiv.org} / \textsc{hal}}} 

\numberwithin{equation}{section}

\newcommand{\Z}{\mathbb{Z}}
\newcommand{\E}{\mathscr{E}}

\newcommand{\ui}{\mathrm{i}}
\newcommand{\ue}{\mathrm{e}}

\newcommand{\eps}{\varepsilon}
\renewcommand{\O}{\mathcal{O}}

\newcommand{\x}{\boldsymbol{x}}
\renewcommand{\k}{\boldsymbol{k}}

\newcommand{\ie}{\emph{i.e.}~}
\newcommand{\eg}{\emph{e.g.}~}

\begin{document}

\title[\Title]{Resonance enhancement by suitably chosen frequency detuning}

\author[D.~Dutykh]{Denys Dutykh}
\address{LAMA, UMR 5127 CNRS, Universit\'e Savoie Mont Blanc, Campus Scientifique, 73376 Le Bourget-du-Lac Cedex, France}
\email{Denys.Dutykh@univ-savoie.fr}
\urladdr{http://www.denys-dutykh.com/}

\author[E.~Tobisch]{Elena Tobisch$^*$}
\address{Institute for Analysis, Johannes Kepler University, Linz, Austria}
\email{Elena.Tobisch@jku.at}
\urladdr{http://www.dynamics-approx.jku.at/lena/}
\thanks{$^*$ Corresponding author}


\begin{titlepage}
\setcounter{page}{1}
\thispagestyle{empty} 
\noindent
{\Large Denys \textsc{Dutykh}}\\
{\it\textcolor{gray}{CNRS--LAMA, University Savoie Mont Blanc, France}}\\[0.02\textheight]
{\Large Elena \textsc{Tobisch}}\\
{\it\textcolor{gray}{Johannes Kepler University, Linz, Austria}}\\[0.16\textheight]

\colorbox{Lightblue}{
  \parbox[t]{1.0\textwidth}{
    \centering\huge\sc
    \vspace*{0.7cm}

    Resonance enhancement by suitably chosen frequency detuning

    \vspace*{0.7cm}
  }
}

\vfill 

\raggedleft     
{\large \plogo} 
\end{titlepage}


\newpage
\maketitle
\thispagestyle{empty}

\begin{abstract}

In this Letter we report new effects of resonance detuning on various dynamical parameters of a generic 3-wave system. Namely, for suitably chosen values of detuning the variation range of amplitudes can be significantly wider than for exact resonance. Moreover, the range of energy variation is not symmetric with respect to the sign of the detuning. Finally, the period of the energy oscillation exhibits  non-monotonic dependency on the magnitude of detuning. These results have important theoretical implications where nonlinear resonance analysis is involved, such as geophysics, plasma physics, fluid dynamics. Numerous practical applications are envisageable \eg in energy harvesting systems.

\bigskip
\noindent \textbf{\keywordsname:} nonlinear resonance; frequency detuning; 3-wave system; resonance enhancement.

\smallskip
\noindent \textbf{MSC:} 

\end{abstract}

\maketitle


\newpage
\tableofcontents
\thispagestyle{empty}
\newpage


\section{Introduction}

Numerous natural phenomena exhibit linear and nonlinear resonances. In many technical cases occurrence of resonance  must to be avoided, the widely known Tacoma Bridge dramatic collapse being an example for this. In other cases, the goal is to approach the state of exact resonance, by reducing resonance detuning, in order to increase the efficiency of a process or device. To give a notion of  linear resonance, we consider a linear oscillator (or pendulum) driven by a small force. We say, that the resonance occurs, if the eigenfrequency $\omega$ of a system coincides with the frequency of the driving force $\Omega$. In this case, for small enough resonance detuning, $|\Omega - \omega| > 0$, the amplitude of the linear oscillator becomes smaller with increasing detuning.

The simplest case of nonlinear resonance is a set of 3 waves $A_j\ue^{\ui(\k_j\x_j - \omega_j t)}$ fulfilling resonance conditions of the following form
\begin{eqnarray}
  \omega_1 \pm \omega_2 \pm \omega_3 &=& 0, \label{eq:resc1}\\
  \k_1 \pm \k_2 \pm \k_3 &=& 0, \label{eq:resc2}
\end{eqnarray}
where $\k_j\in\Z^2$, $\omega_j = \omega(\k_j)$ are the wave vectors and frequencies respectively. Drawn from these resonance conditions, resonance detuning in the nonlinear case can be defined in a number of ways, \eg as a frequency detuning or phase detuning.

In the physical literature, for frequency detuning defined as
\begin{equation*}
  \omega_1\ +\ \omega_2\ -\ \omega_3\ =\ \widetilde{\Delta\omega}\ \ll\ \min\limits_{j=1,2,3}\{\omega_j\},
\end{equation*}
the assumption prevails, that bigger detuning  results in  smaller variation of amplitudes, \eg \cite{Craik1988}. Quite consistently the study of phase detuning in \eqref{eq:resc2} in a 3-wave system  demonstrated that the variation range of wave amplitudes becomes smaller with growing dynamical phase \cite{Bustamante2009a}.

In this Letter we study the effects of frequency detuning in \eqref{eq:resc1} by means of the numerical simulation. As an example we take a resonant triad of atmospheric planetary waves from \cite{Kartashova2007}. Our numerical simulations show that the detuned system behaviour was correctly understood for $|\widetilde{\Delta\omega}| \gg \delta > 0$. However, there exists a transitional range of the detuning magnitude $\widetilde{\Delta\omega}$ where the dynamics of this system is much more complicated. As our results depend only on the general form of the dynamical system given below, they are applicable to a wide class of physical systems.

\section{Model equations}

The dynamical system for  detuned resonance of three complex-valued amplitudes $A_i$, $i=1,2,3$ reads
\begin{eqnarray*}\label{3-comp}
  N_1\dot{A_1} &=& -2\ui Z(N_2 - N_3)A_2^* A_3\ue^{-\ui\Delta\omega T}, \\
  N_2\dot{A_2} &=& -2\ui Z(N_3 - N_1)A_1^* A_3\ue^{-\ui\Delta\omega T}, \\
  N_3\dot{A_3} &=& 2\ui Z(N_1 - N_2)A_1 A_2\ue^{\ui\Delta\omega T},
\end{eqnarray*}
(and their complex conjugate equations) where the dot denotes  differentiation with respect to the slow time $T = t/\eps$, $\Delta\omega := \widetilde{\Delta\omega}/\eps$, $\eps$ being a small parameter. The dynamical system for exact resonance is obtained by setting $\Delta\omega\equiv 0$. It can be rewritten in amplitude/phase variables as:
\begin{eqnarray}\label{eq:ds1}
  N_1\dot{C_1} &=& -2Z(N_2 - N_3)C_2 C_3\sin\psi, \\
  N_2\dot{C_2} &=& -2Z(N_3 - N_1)C_1 C_3\sin\psi, \\
  N_3\dot{C_3} &=& -2Z(N_1 - N_2)C_1 C_2\sin\psi, \\
  \dot{\psi} &=& \Delta\omega - 2ZC_1 C_2 C_3\Bigl(\frac{N_2 - N_3}{N_1}C_1^{-2} + \frac{N_3 - N_1}{N_2}C_2^{-2} + \frac{N_1 - N_2}{N_3}C_3^{-2}\Bigr)\cos\psi, \label{eq:ds4}
\end{eqnarray}
where $C_i(T) = |A_i(T)|$ is the real amplitude, $\psi := \theta_1 + \theta_2 - \theta_3$ is the dynamical phase and $\theta_i(T) = \arg A_i(T)$. In what follows we will focus on the evolution of the energy of the high-frequency mode  $\E_3(T)$.

\section{Amplitudes}

\begin{figure}
  \centering
  \includegraphics[width=0.79\textwidth]{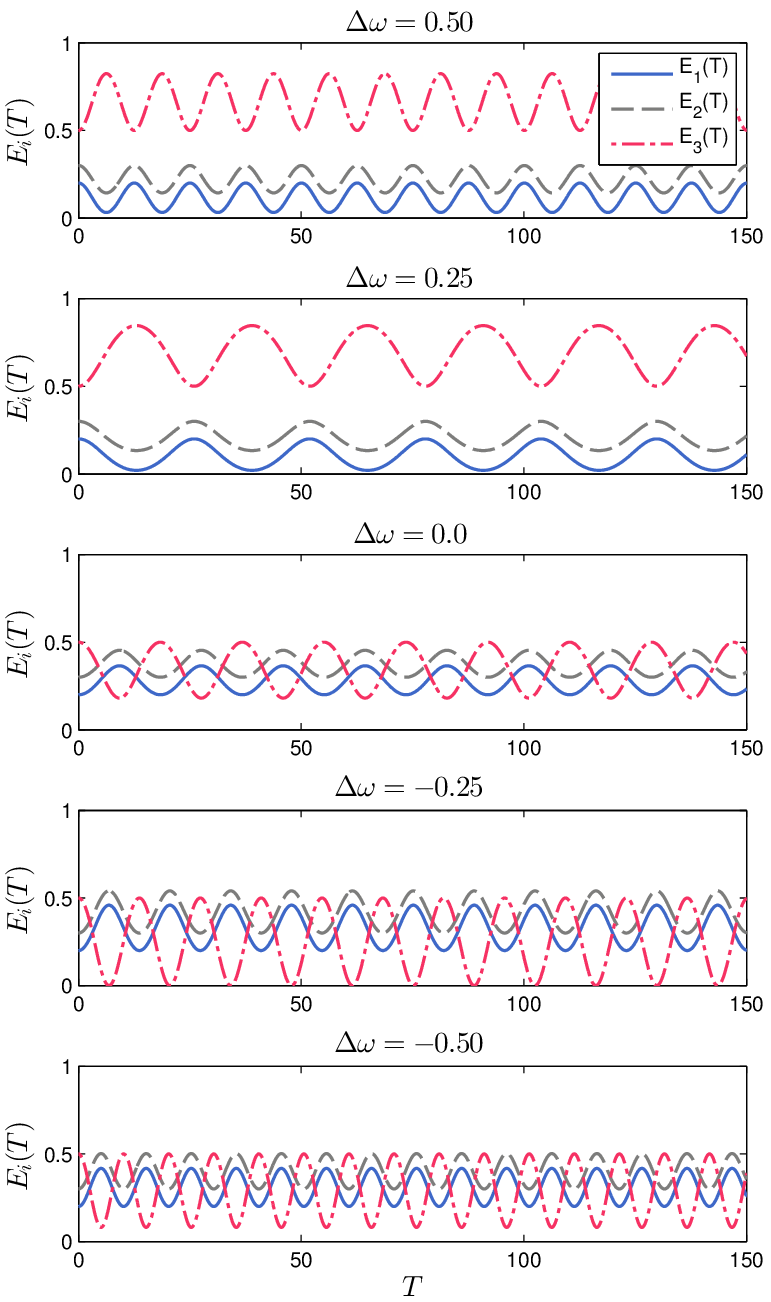}
  \caption{\small\em Energy evolution in the triad given in Table~\ref{tab:params}, for different values of the detuning $\Delta\omega$.}
  \label{fig:evolution}
\end{figure}

In Fig.~\ref{fig:evolution} we show the energy evolution in the resonant triad given in Table~\ref{tab:params} for several values of the frequency detuning $\Delta\omega = \widetilde{\Delta\omega}/\eps \in [-\frac12, \frac12]$; \eg in geophysical applications $\eps\sim \O(10^{-2})$. From these graphs it can be seen that the period $\tau$ and the range of the energy variation, defined as
\begin{equation}\label{eq:energy}
  \Delta\E(\Delta\omega)\ :=\ \frac12\bigl(\max\limits_{t}{\E}\ -\ \min\limits_{t}{\E}\bigr),
\end{equation}
are \emph{non-monotonic} functions of the detuning $\Delta\omega$.

\begin{table}
  \centering
  \begin{tabular}{c|c}
  \hline\hline
  \textit{Parameter} & \textit{Value} \\
  \hline\hline
  Resonant  wavevectors, $[m_j,n_j]$ & $[4,12]$, $[5,14]$, $[9,13]$ \\
  Resonant  frequencies, $2m_j/n_j(n_j+1)$ & $0.0513$, $0.0476$, $0.0989$ \\
  Resonant triad parameters, $N_j$ & $156$, $210$, $182$ \\
  Interaction coefficient, $Z$ & $7.82$ \\
  Initial energy distribution (a), \% & $20$\%, $30$\%, $50$\% \\
  Initial energy distribution (b), \% & $40$\%, $40$\%, $20$\% \\
  Initial dynamical phase, $\psi$ & 0.0 \\
  \hline\hline
  \end{tabular}
  \bigskip
  \caption{Physical parameters used in numerical simulations.}
  \label{tab:params}
\end{table}

\begin{figure}
  \centering
  \includegraphics[width=0.69\textwidth]{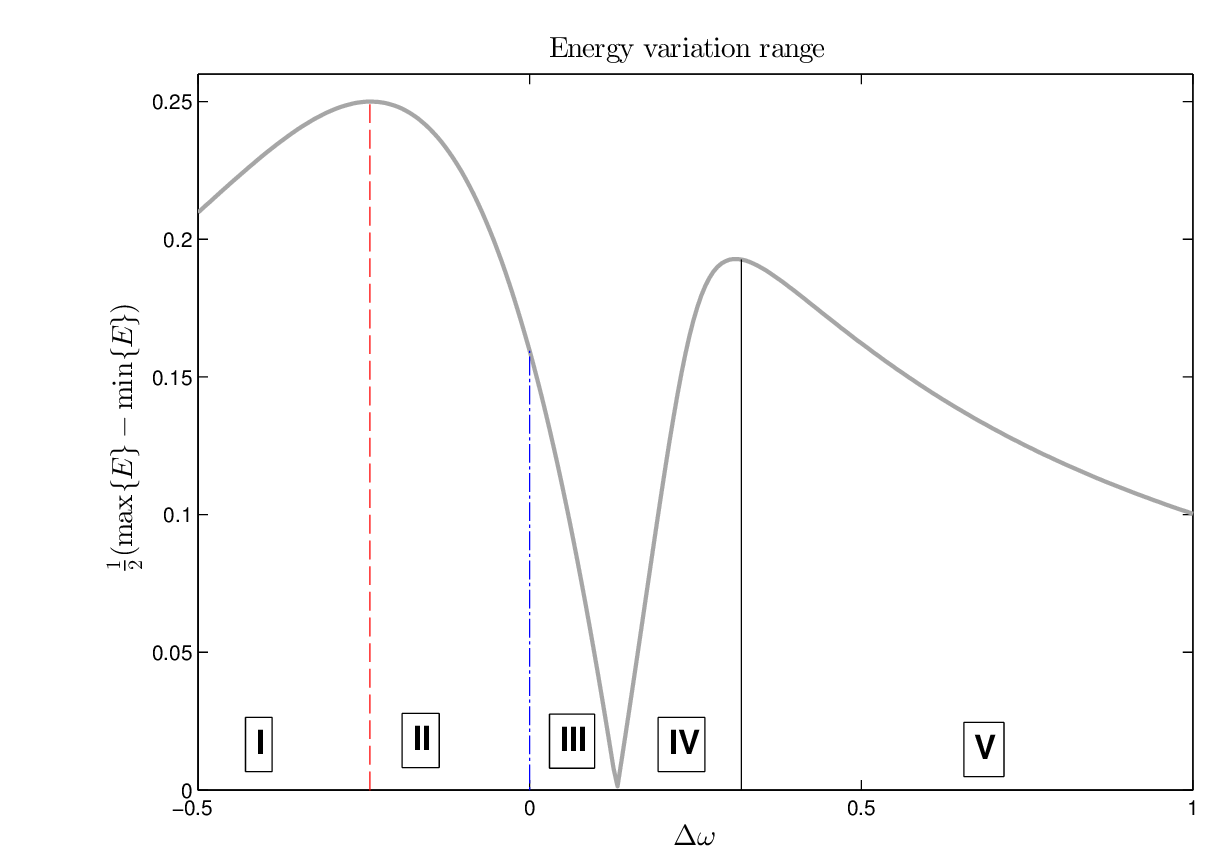}
  \caption{\small\em Typical dependency of the energy variation range $\Delta\E$ on the frequency detuning $\Delta\omega$ for the case when the high frequency mode $\omega_3$ has the maximal energy (initial condition (a)). The vertical red dashed line shows the location of $\Delta\omega_{\max}^{(1)}$, while the vertical black solid line shows the location of $\Delta\omega_{\max}^{(2)}$. Finally, the blue dash-dotted line shows the amplitude obtained the exact resonance.}
  \label{fig:ampl}
\end{figure}

A graph showing the characteristics of the dependency of the energy variation range $\Delta\E(\Delta\omega)$ on the frequency detuning $\Delta\omega$ is shown in Fig.~\ref{fig:ampl}. This particular curve was computed for parameters given in Tab.~\ref{tab:params}. The graph can conveniently be divided into the five regions which are separated by particular values of the  frequency detuning $\Delta\omega$: $\Delta\omega_{\max}^{(1,2)}$ correspond to local maxima, $\Delta\omega_{\mathrm{st}}$ is the position of the local minimum, and $\Delta\omega = 0$ corresponds to exact resonance. So, the regions are:

\begin{description}
  \item[(I)] $\Delta\omega\in \bigl(-\infty, \Delta\omega_{\max}^{(1)}\bigr]$;
  \item[(II)] $\Delta\omega\in \bigl(\Delta\omega_{\max}^{(1)}, \min\{0, \Delta\omega_{\mathrm{st}}\}\bigr]$;
  \item[(III)] $\Delta\omega\in \bigl(\min\{0, \Delta\omega_{\mathrm{st}}\}, \max\{0, \Delta\omega_{\mathrm{st}}\}\bigr]$;
  \item[(IV)] $\Delta\omega\in \bigl(\Delta\omega_{\mathrm{st}}, \Delta\omega_{\max}^{(2)}\bigr]$;
  \item[(V)] $\Delta\omega\in \bigl(\Delta\omega_{\max}^{(2)}, +\infty\bigr)$.
\end{description}

The  reason to regard these regions separately is that the main characteristics (\ie energy variation $\Delta\E$, energy oscillation period $\tau$ and the phase variation $\Delta\psi$) behave differently in each region. Our findings are summarized in Table~\ref{tab:behave}, where all the quantities $\E$, $\tau$ and $\psi$ are followed by $\pm$ sign denoting the their variation in the region ($+$: increase, $-$: decrease). The first column corresponds to the direction of increasing values of $\Delta\omega\in (-\infty\rightarrow +\infty)$, while the second column corresponds to the opposite direction $\Delta\omega\in(-\infty\leftarrow +\infty)$.

\begin{table}
  \centering
  \begin{tabular}{c|c|c}
  \hline\hline
  \textit{Region/Range} & $\longrightarrow$ & $\longleftarrow$ \\
  \hline\hline
  (I) & $\Delta\E+$, $\tau+$, $\Delta\psi-$ & $\Delta\E-$, $\tau-$, $\Delta\psi+$ \\
  (II) & $\Delta\E-$, $\tau+$, $\Delta\psi-$ & $\Delta\E+$, $\tau-$, $\Delta\psi+$ \\
  (III) & $\Delta\E-$, $\tau+$, $\Delta\psi-$ & $\Delta\E+$, $\tau-$, $\Delta\psi+$ \\
  (IV) & $\Delta\E+$, $\tau-$, $\Delta\psi+$ & $\Delta\E-$, $\tau-$, $\Delta\psi-$ \\
  (V) & $\Delta\E-$, $\tau-$, $\Delta\psi+$ & $\Delta\E+$, $\tau+$, $\Delta\psi-$ \\
  \hline\hline
  \end{tabular}
  \bigskip
  \caption{\small\em Behaviour of physical parameters $\Delta\E$, $\tau$ and $\psi$ in different regions.}
  \label{tab:behave}
\end{table}

\begin{figure}
  \centering
  \includegraphics[width=0.60\textwidth]{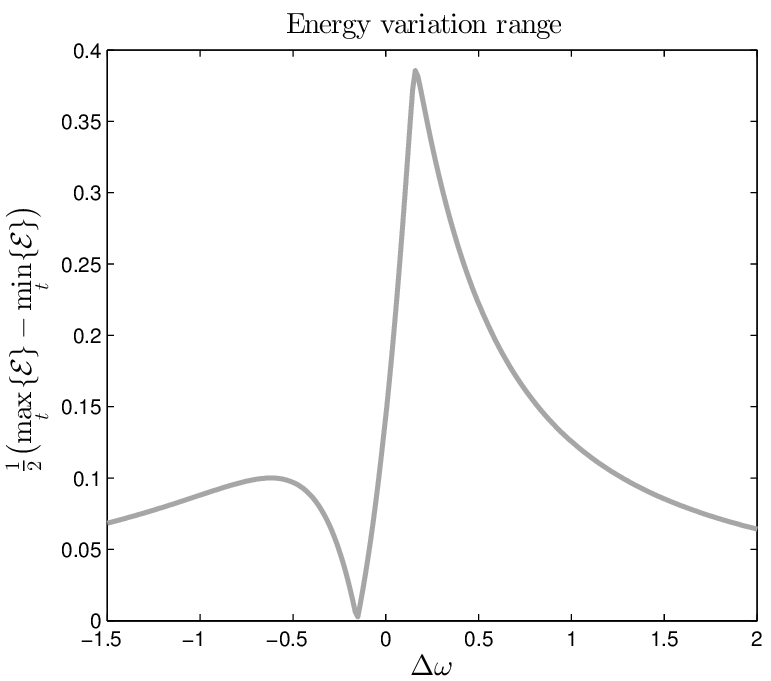}
  \caption{\small\em Typical dependence of the energy variation range $\Delta\E$ on the frequency detuning $\Delta\omega$ for the case when the high frequency mode $\omega_3$ has the lowest energy (initial condition (b)).}
  \label{fig:right}
\end{figure}

The most interesting observation is, that the energy variation range during the system evolution can be significantly larger for a suitable choice of the detuning $\Delta\omega \neq 0$ compared to the exact resonance case $\Delta\omega = 0$.
There are two values of  which provide significant amplifications to $\Delta\E$. On Fig.~\ref{fig:ampl} the global maximum is located on the left of $\Delta\omega_{\mathrm{st}}$, while on Fig.~\ref{fig:right} it is on the right of $\Delta\omega_{\mathrm{st}}$. These two cases differ only by the initial energy distribution among the triad modes (see Tab.~\ref{tab:params}, initial conditions (a) \& (b)).

Similar computations have been performed for other resonant triads and the qualitative behavior of the energy variation has always been similar to Figs.~\ref{fig:ampl} \& \ref{fig:right}. Namely, the global maximum is located on the left of $\Delta\omega_{\mathrm{st}}$ when the high frequency mode $\omega_3$ contains initially most of the energy, and to the right of $\Delta\omega_{\mathrm{st}}$ in the opposite case.

It is important to stress that the energy variation $\Delta\E(\Delta\omega)$ at the global maximum $\Delta\omega_{\max}^{(\mathrm{g})}$ is always significantly higher than at the point of exact resonance, \ie $\Delta\E(\Delta\omega_{\max}^{(\mathrm{g})}) > \Delta\E(0)$. The highest ratio $\Delta\E(\Delta\omega)/\Delta\E(0)$ is attained when the local minimum $\Delta\omega_{\mathrm{st}}$ coincides with the point of exact resonance. In this case we can find a $\Delta\omega$ that the amplification $\Delta\E(\Delta\omega)/\Delta\E(0)$ is of at least one order of magnitude. A simple phase space analysis allows to locate the local minimum.

\section{Phase space analysis}

\begin{figure}
  \centering
  \subfigure[$\ \Delta\omega = 0.1$]{
  \includegraphics[width=0.47\textwidth]{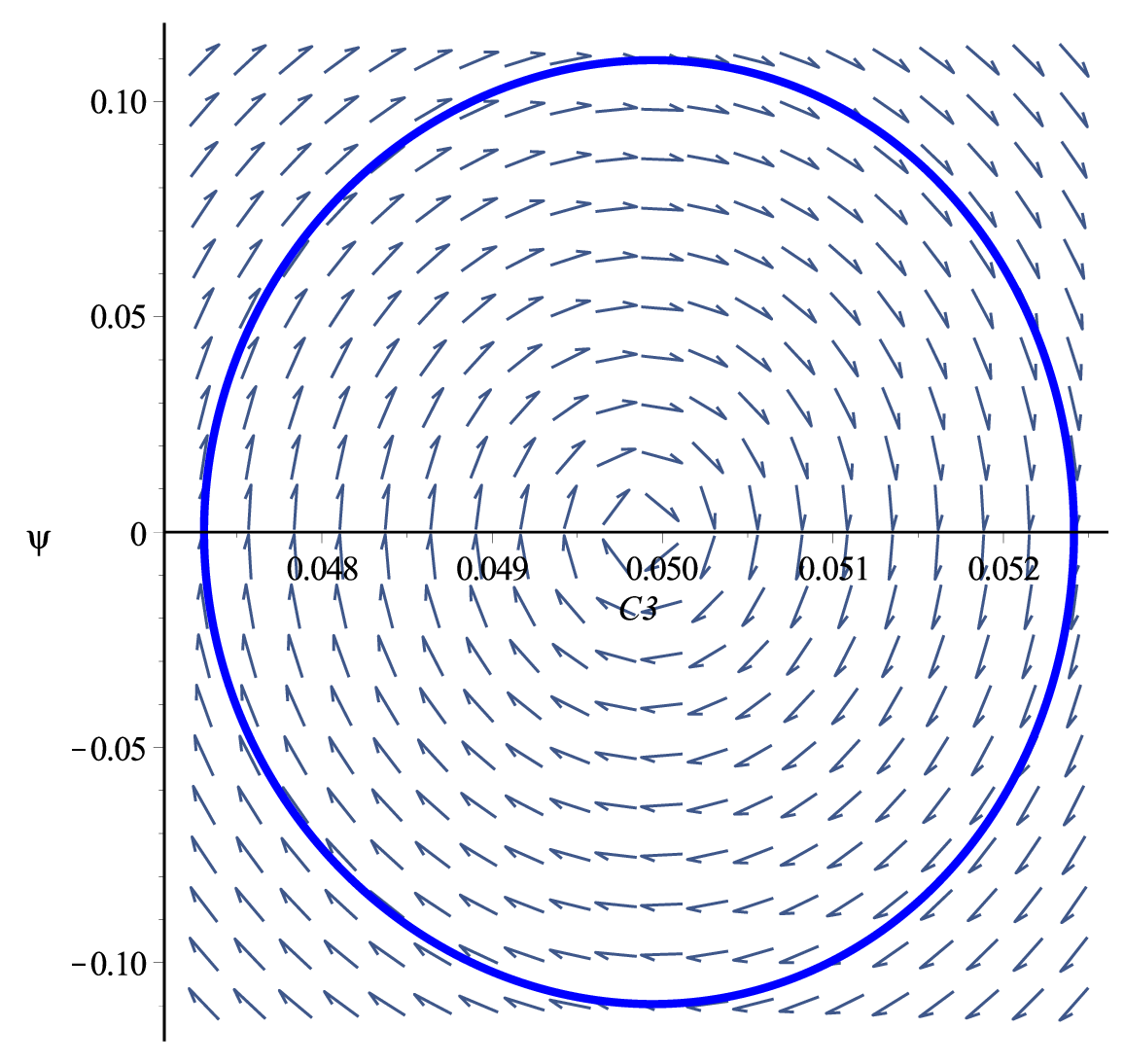}
  }
  \subfigure[$\ \Delta\omega = 0.31$]{
  \includegraphics[width=0.47\textwidth]{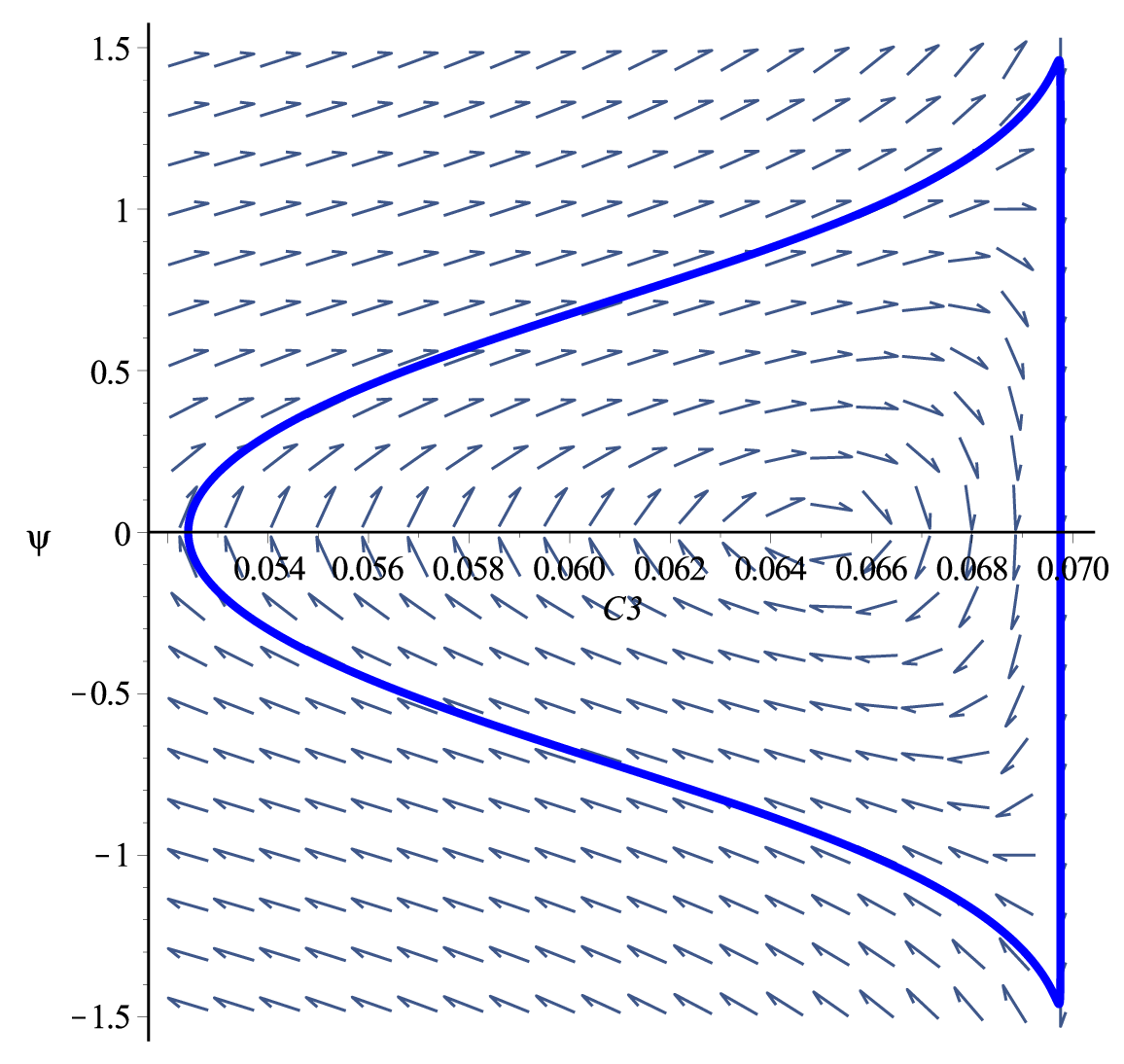}
  }
  \subfigure[$\ \Delta\omega = 0.315$]{
  \includegraphics[width=0.77\textwidth]{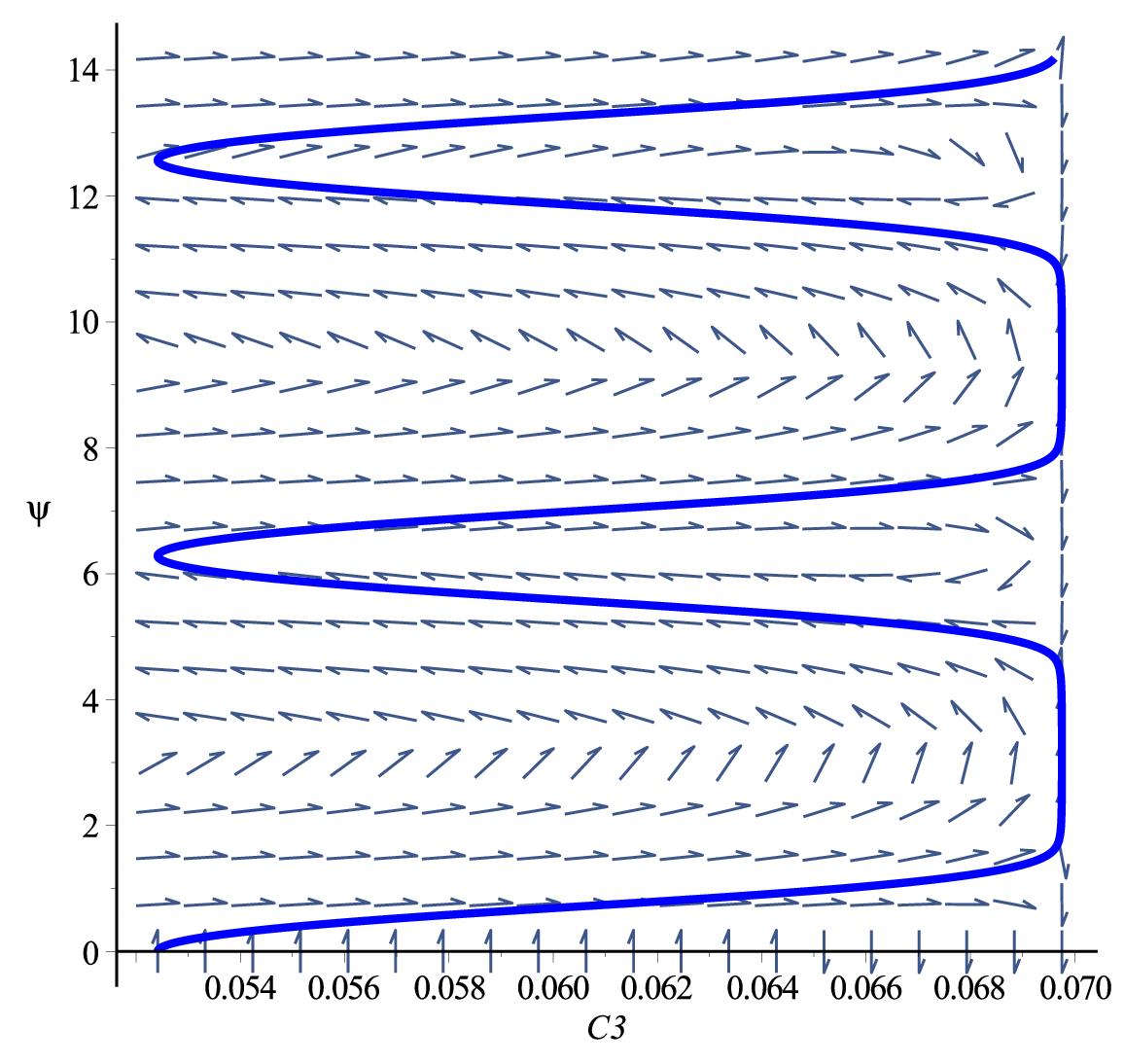}
  }
  \caption{\small\em Phase portraits of the dynamical system \eqref{eq:ds1} -- \eqref{eq:ds4} in $(C_3, \psi)$ variables for the triad from the Tab.~\ref{tab:params}(a). Positive increasing detuning.}
  \label{fig:phase1}
\end{figure}

\begin{figure}
  \centering
  \subfigure[$\ \Delta\omega = -0.1$]{
  \includegraphics[width=0.47\textwidth]{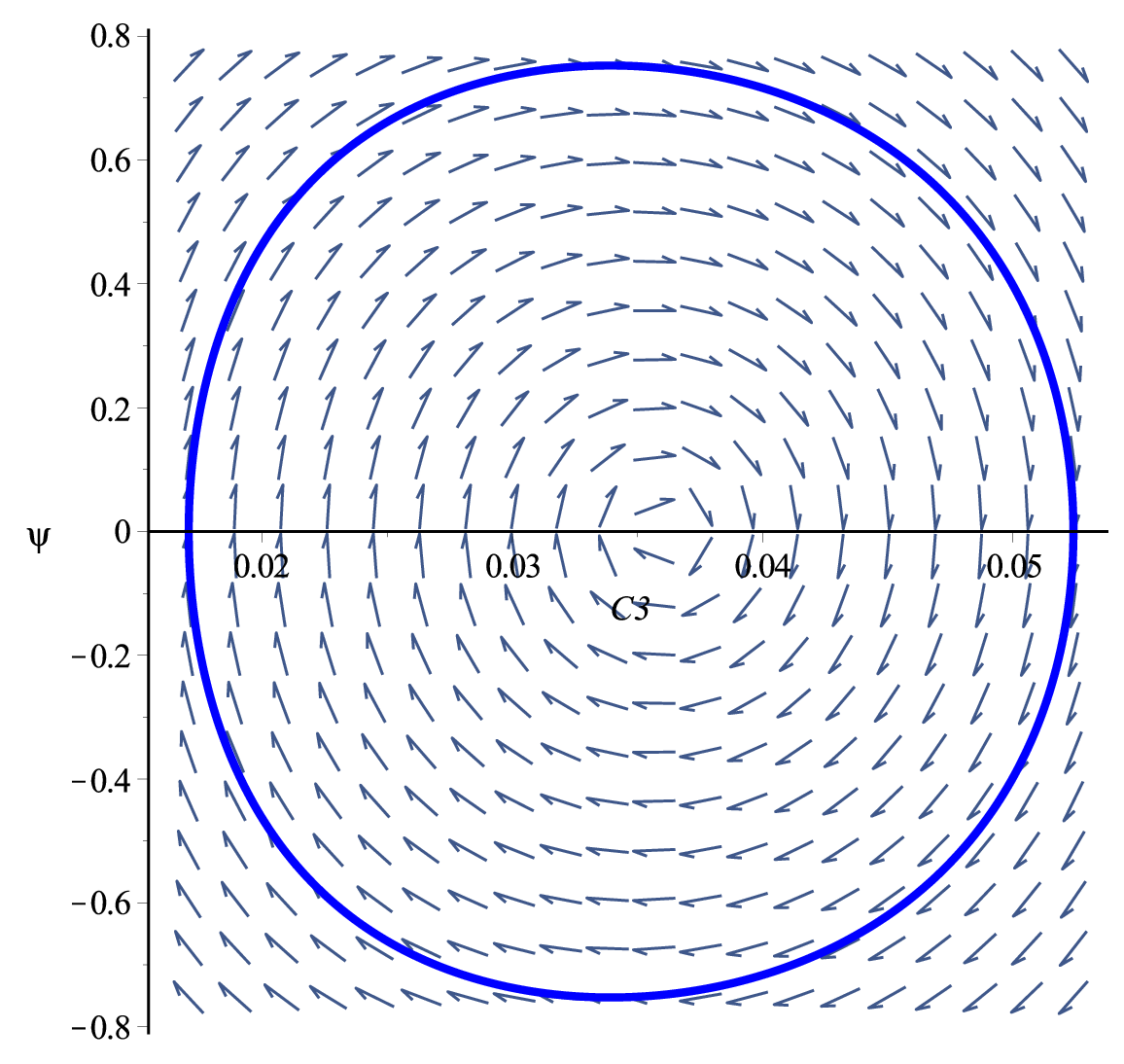}
  }
  \subfigure[$\ \Delta\omega = -0.235$]{
  \includegraphics[width=0.47\textwidth]{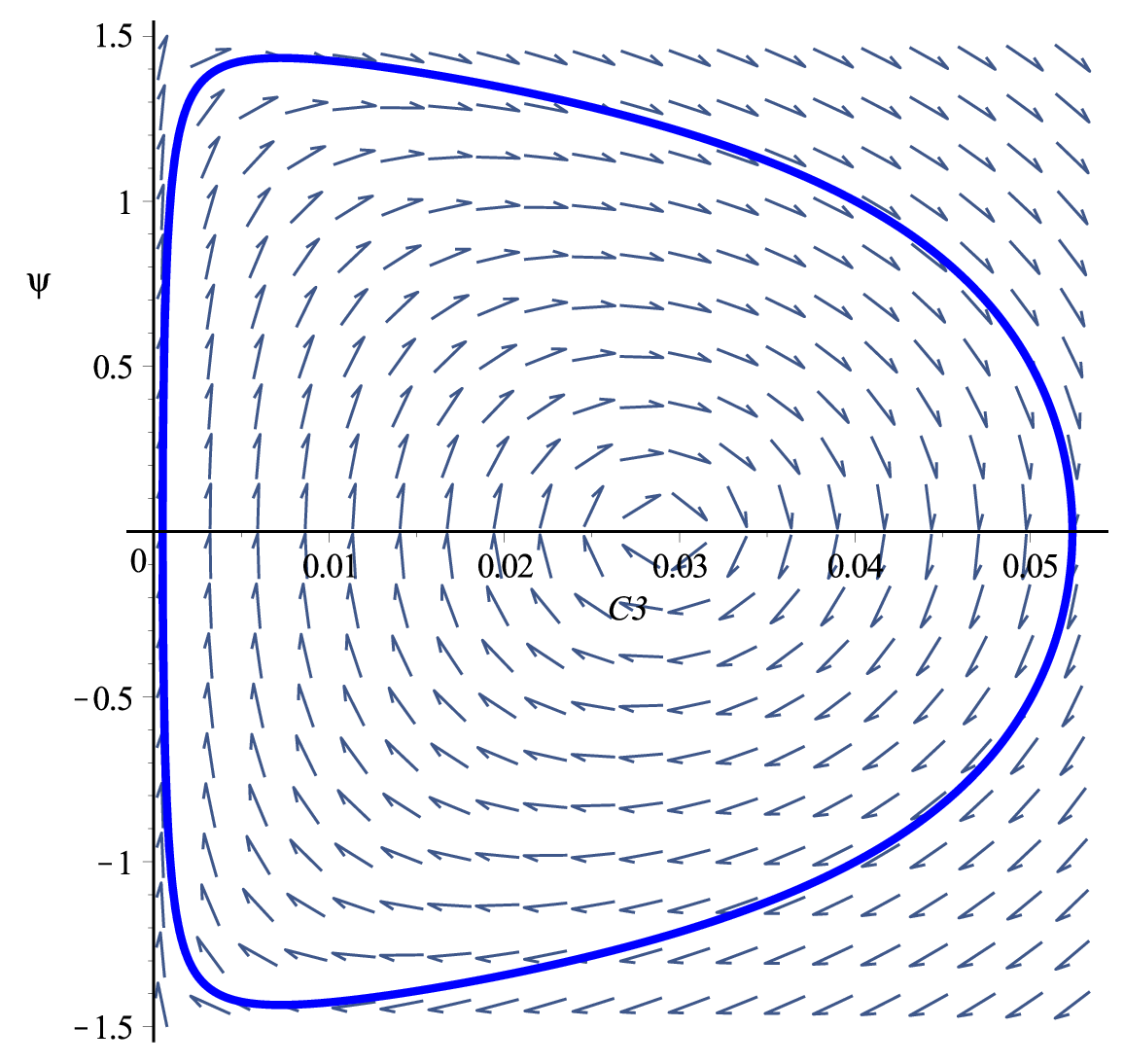}
  }
  \subfigure[$\ \Delta\omega = -0.25$]{
  \includegraphics[width=0.77\textwidth]{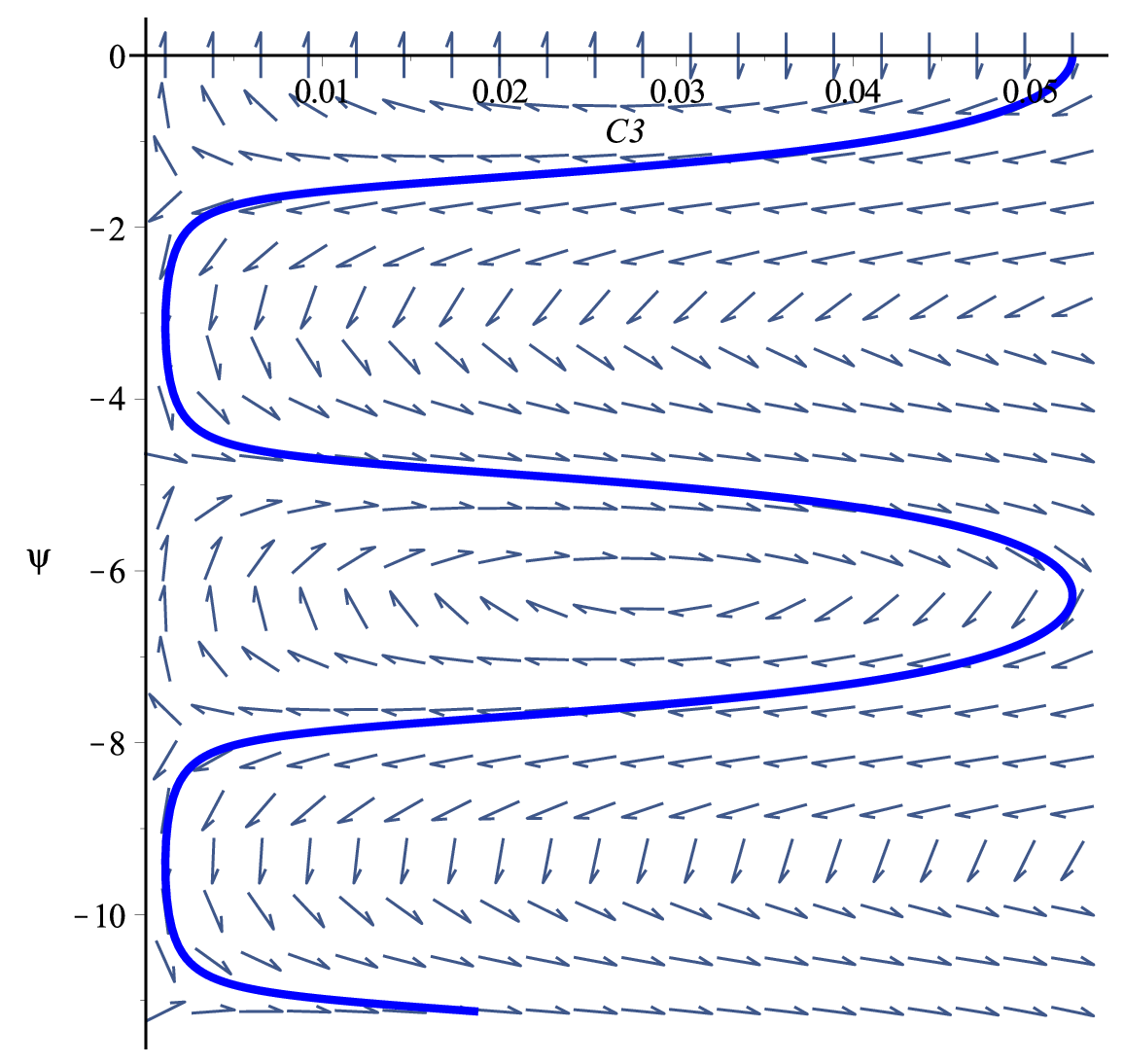}
  }
  \caption{\small\em Phase portraits of the dynamical system \eqref{eq:ds1} -- \eqref{eq:ds4} in $(C_3, \psi)$ variables for the triad from the Tab.~\ref{tab:params}(a). Negative decreasing detuning.}
  \label{fig:phase2}
\end{figure}

\begin{figure}
  \centering
  \includegraphics[width=0.79\textwidth]{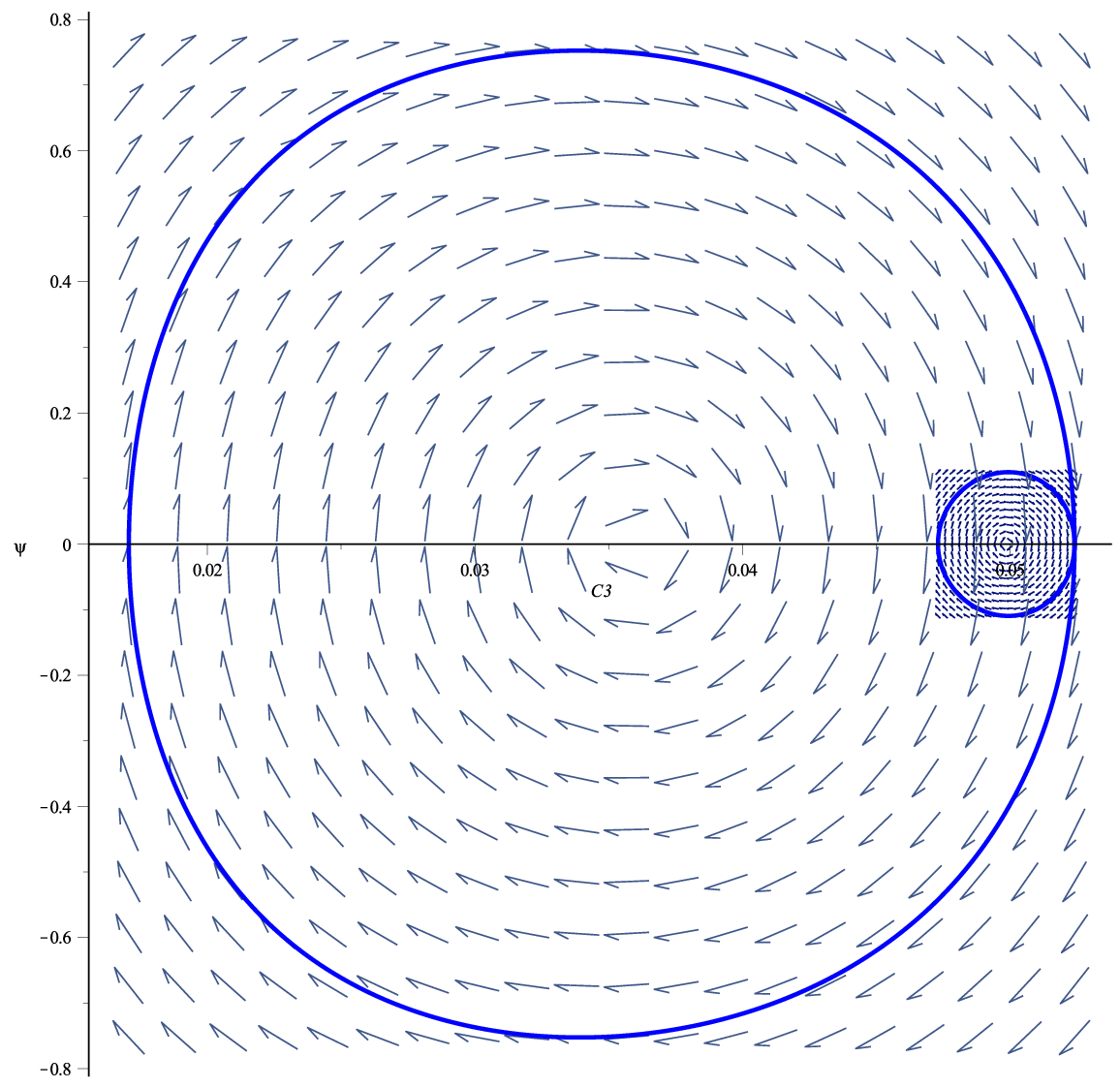}
  \caption{\small\em Simultaneous plot of two phase portraits and integral curves for the detunings $\Delta\omega = \pm 0.1$ shown at Figs.~\ref{fig:phase1}(a) \& \ref{fig:phase2}(a) correspondingly.}
  \label{fig:together}
\end{figure}

On Figs.~\ref{fig:phase1} -- \ref{fig:together} we depict the typical phase portraits of the dynamical system \eqref{eq:ds1} -- \eqref{eq:ds4} in phase-amplitude variables. For illustration we choose the triad given in Tab.~\ref{tab:params} with the initial energy distribution (a). In these pictures we represent the high-frequency mode $C_3$ on the horizontal axis, while the dynamical phase $\psi$ is on the vertical.

The main finding is a pronounced asymmetry between the phase portraits for positive and negative values of the detuning $\Delta\omega$: the position of the stationary point right on the horizontal axis $C_3$ (there are other stationary points for $\psi \neq 0$) is very different; the shape of the periodic orbit differs (see Figs.~\ref{fig:phase1}(b) \& \ref{fig:phase2}(b)); the transition from closed to snake-like integral curves takes place for different values of $|\Delta\omega|$, \eg $\approx 0.31$ on Fig.~\ref{fig:phase1}(c) and $\approx -0.24$ on Fig.~\ref{fig:phase2}(c); the shape of the integral curves is different; the phase portraits look alike, but differ in size by one order of magnitude. In order to demonstrate how big this difference is for opposite values of $\Delta\omega$ we depicted on the same Fig.~\ref{fig:together} the periodic cycles from Figs.~\ref{fig:phase1}(a) \& \ref{fig:phase2}(a).

A simple phase space analysis reveals the reason for the presence of a local minimum of $\Delta\E$ in Figs.~\ref{fig:ampl}~\& \ref{fig:right}. Indeed, it can happen that the initial conditions coincide with the system equilibrium point, which depends on $\Delta\omega_{\mathrm{st}}$.

\section{Conclusions}

It was demonstrated that the introduction of frequency detuning significantly enriches the dynamics of a 3-wave resonance system. Moreover, the effects of detuning are highly nonlinear and highly non-monotonic with respect to the detuning parameter. The main findings of this study are outlined hereinbelow:
\begin{itemize}
  \item The range of values of frequency detuning $\Delta\omega$ can most conveniently be divided into five regions of different behaviour, not all of them present in any case. The behavior of the main parameters of system \eqref{eq:ds1} -- \eqref{eq:ds4} over those five regions is summarized in Tab.~\ref{tab:behave}.

  \item The amplitude of energy variation  \eqref{eq:energy} in a triad with suitably chosen detuning ($\Delta\omega \neq 0$) can be significantly higher than in the case of exact resonance, \ie $\Delta\omega \equiv 0$. The maximal amplification as compared to exact resonance is attained when $\Delta\omega_{\mathrm{st}}$ coincides with the point of exact resonance. In this case one of the zones (III) or (IV) disappears.

  \item The phase portraits along with the shape and size of the periodic cycles are substantially different for $\Delta\omega > 0$ and $\Delta\omega < 0$. This means that any complete analysis of detuned resonance must include both positive and negative values of the detuning parameter $\Delta\omega$.
\end{itemize}

The notion of \emph{resonance enhancement via frequency detuning}, as the title of this Letter says, contradicts  what we would expect from our physical intuition. However, there exists a simple qualitative explanation for that. Indeed, our intuition comes from a \emph{linear pendulum}
\begin{equation}\label{lin-pend}
  \ddot{x}\ +\ \omega^2x\ =\ 0
\end{equation}
taken usually as a model for a linear wave, and a resonance is regarded due to an action of an external force.

On the other hand, the dynamical system for 3-wave resonance can be transformed into the Mathieu equation which describes a particular case of the motion of an \emph{elastic pendulum}
\begin{equation}\label{eq:Mathieu}
  \ddot{x}\ +\ [\omega_{\mathrm{pen}}^2\ -\ \lambda \cos(\omega_{\mathrm{spr}})]x\ =\ 0,
\end{equation}
where $\omega_{\mathrm{pen}}$ and $\omega_{\mathrm{str}}$ are frequencies of  pendulum- and spring-like motions, \cite{Kartashova2010}. Regarding resonance detuning $\widetilde{\Delta\omega}$ as a frequency of an external force for \eqref{eq:Mathieu}, our findings can be understood in the following way. The detuned 3-wave system has the maximal range of the energy amplitudes variation when the elastic pendulum interacts resonantly with the external forcing. Detailed study of this effect is upcoming, using the approach developed in \cite{Gitterman2010} for an elastic pendulum subject to the external force.


\subsection*{Acknowledgments}
\addcontentsline{toc}{subsection}{Acknowledgments}

This research has been supported by the Austrian Science Foundation (FWF) under projects P22943-N18 and P24671.


\addcontentsline{toc}{section}{References}
\bibliographystyle{abbrv}
\bibliography{biblio}

\end{document}